\documentclass[preprint,eqsecnum,preprintnumbers,nofootinbib,byrevtex,prd,aps,showpacs,showkeys,groupedaddress,
floatfix]{revtex4}
\usepackage{bm}
\usepackage{graphics}
\usepackage{graphicx}
\usepackage{epsfig}
\usepackage{amssymb}
\usepackage{amsmath}
\begin{document}

\title{Analytical solutions of the  Klein-Fock-Gordon
equation  with the Manning-Rosen  potential plus a Ring-Shaped like
potential} 
\author{A.~I.~Ahmadov$^{1,3}$~\footnote{E-mail:
ahmadovazar@yahoo.com}}
\author{C.~Aydin$^{2,4}$~\footnote{E-mail: coskun@ktu.edu.tr}}
\author{O.~Uzun$^{2}$~\footnote{E-mail: $\mbox{oguzhan}_-\mbox{deu@hotmail.com}$}}

\affiliation {$^{1}$ Theory Division, Physics Department, CERN, CH-1211 Ceneva 23, Switzerland \\
$^{2}$ Department of Physics, Karadeniz Technical
University, 61080, Trabzon, Turkey \\
$^{3}$ Institute for  Physical Problems, Baku State University, Z.
Khalilov st. 23, AZ-1148, Baku, Azerbaijan\\
$^{4}$ Department of Physics, University of Surrey, Guildford Surrey
GU2 7XH, UK}

\date{}

\begin{abstract}
In this work, on the condition that scalar potential is equal to
vector potential, the bound state solutions of the Klein-Fock-Gordon
equation of the Manning-Rosen plus ring-shaped like potential are
obtained by Nikiforov-Uvarov method. The energy levels are worked
out and the corresponding normalized eigenfunctions are obtained in
terms of orthogonal polynomials for arbitrary $l$ states. The
conclusion also contain central Manning-Rosen, central and
non-central Hulth\'en potential.
\end{abstract}

 \maketitle

\section{INTRODUCTION.}
Since the early years of quantum mechanics (QM) the study of exactly
solvable problems for some special potentials of physical interest
has attracted much attention in theoretical physics. Obtaining the
analytical  solutions of the Klein-Fock-Gordon(KFG), Dirac and other
wave equations one of the interesting problems in high energy and
nuclear physics. These wave  equations  are frequently used to
describe the particle dynamics in relativistic quantum mechanics.
Already long time in literature, a great deal of effort has been
spent to solve these relativistic wave equations for different
potentials which also include mixing potentials.

The description of phenomena at high energies requires the
investigation of relativistic wave equations, which are invariant
under Lorentz transformation, to give correction for nonrelativistic
QM\cite{Greiner,Bagrov}. If we consider the case where the
interaction potential is not enough to create particle-antiparticle
pairs, we can apply the KFG equation to the treatment of a zero-spin
particle and apply the Dirac equation to that of a $1/2$-spin
particle.  A particle is moving in a strong potential field, the
relativistic effect must be considered. This effect  gives the
correction for nonrelativistic QM. Taking the relativistic effects
into account, a particle including mixing potential should be
described by the Klein-Fock-Gordon and Dirac equations.

In Refs. \cite{adame1,adame2,rao1,rao2, villalba,ma,
samsonov,mustafa1,mustafa2,mustafa3, alihaidari1, alihaidari2,
alihaidari3,
alihaidari4,simsek,hu,hou1,hou2,guo,chen1,chen2,chen3,chen4,chen5,chen6,qiang1,qiang2,qiang3,qiang4,qiang5,boztosun,Badalov2,dong1,diao}
analytical solutions of the KFG and Dirac equations are widely
studied.

Many methods were developed and has been used successfully in
solving the Schr\"{o}dinger, Dirac and KFG wave equations in the
presence of some well-known potentials. In
Refs.\cite{hu,hou1,hou2,guo,chen1,chen2,chen3,chen4,chen5,chen6,qiang1,qiang2,qiang3,qiang4,qiang5,boztosun,dong1,diao}
some authors have assumed that the scalar potential is equal to the
vector potential and using NU \cite{Nikiforov} method obtained bound
states of the KFG and Dirac equation with some typical potential
fields.

The noncentral potentials are needed to obtain better results than
central potentials about the dynamical properties of the molecular
structures and interactions. Some authors added ring-shaped
potentials to certain potentials, for example Coulomb, Hulth\'en and
Manning–Rosen potentials to obtain noncentral potentials.

Many works show the power and simplicity of NU method in solving
central and noncentral potentials, for example
Ref.\cite{Badalov1,Badalov2,Badalov3,Ahmadov}. This method is based
on solving the second order linear differential equation by reducing
to a generalized equation of hypergeometric type which is a second
order homogeneous differential equation with polynomials
coefficients of degree not exceeding the corresponding order of
differentiation.

It would be interesting and important to study the relativistic
bound states of the arbitrary $l$-wave KFG equation with
Manning-Rosen potential plus a ring-shaped like potential, since it
has been extensively used to describe the bound and continuum states
of the interacting systems. The central Manning-Rosen \cite{Manning}
potential is defined by
\begin{equation}
V(r,\theta)=\frac{1}{kb^2}\left[\frac{\alpha(\alpha-1)exp(-2r/b)}{(1-exp(-r/b))^2}-\frac{Aexp(-r/b)}{(1-exp(-r/b)}\right],
k=2M/\hbar^2,
\end{equation}
where $A$ and $\alpha$ are dimensionless parameters, but the
screening parameter $b$, determines the potential range, has
dimension of length.

This potential is used as a mathematical model in the description of
diatomic molecular vibrations and it constitutes a convenient model
for other physical situations. It is known that for this potential
the KFG equation can be solved exactly using suitable approximation
scheme to deal with the centrifugal term~\cite{Dong}.

The potential which we used in this work
\begin{equation}
V(r,\theta)=\frac{1}{k}\left[\frac{\alpha(\alpha-1)exp(-2r/b)}{b^2
(1-exp(-r/b))^2}-\frac{Aexp(-r/b)}{b^2
(1-exp(-r/b)}+\frac{\beta'}{r^2sin^2\theta}+\frac{\beta
cos\theta}{r^2sin^2\theta}\right],
\end{equation}
is obtained by adding  a ring-shaped like potential term.

Ring-shaped like potentials is usually  used in quantum chemistry
for describing the ring shaped organic molecules such as benzene and
in nuclear physics for investigation the interaction between
deformed pair of nucleus and spin-orbit coupling for the motion of
the particle in the potential fields.

From the point of view of theoretical end experimental physics
Manning-Rosen plus ring-shaped like potential more informative
relative Manning-Rosen potential.

By taking into account these point the solution of the KFG equation
for Manning-Rosen plus ring-shaped like potentials present a great
interest both theoretical and experimental studies.

Here we present the analytical solutions of the KFG equation with
equal scalar and vector Manning-Rosen plus ring-shaped potential.

The remainder of this paper is organized as follows. In Section II,
we provide KFG equation within Manning-Rosen plus a ring-shaped like
potential. In Section III, we present full details of bound state
solution of the radial KFG equation by NU method. In Section IV, we
present the solution of angle-dependent part of the KFG. Finally, we
summarize our results and present our conclusions in Section V.

\section{THE KLEIN-FOCK-GORDON EQUATION WITH THE MANNING-ROSEN POTENTIAL PLUS A RING-SHAPED LIKE POTENTIAL.}

Since KFG equation contains two objects; the four-vector linear
momentum operator and the scalar rest mass, one can introduce two
different potentials in this equation. The first is a vector
potential (V), introduced via minimal coupling and the second is a
scalar potential (S) introduced via scalar coupling \cite{Greiner}.
They allows us to introduce two types of potential coupling which
are the four vector potential (V) and the space-time scalar
potential (S).

The KFG equation with scalar potential $S(r,\theta)$ and vector
potential $V(r,\theta)$ can be written in the following form in
natural units ($\hbar=c=1$)
\begin{equation}
  [-\nabla ^2 +(M+S(r,\theta))^2]\psi (r,\theta,\phi)=[E-V(r,\theta)]^2\psi
  (r,\theta,\phi),
 \end{equation}
where $E$ is the relativistic energy of the system and $M$ denotes
the rest mass of a scalar particle.

Here, we consider the case when the scalar potential and vector
potential are equal to the Manning-Rosen plus ring-shaped potential
as done in \cite{Egrifes} . By taking the wave function of the form

\begin{equation}
\psi (r,\theta,\phi)=\frac {\chi (r)}{r}\Theta
(\theta)e^{im\phi},~~m=0,\pm 1,\pm 2,\pm 3 ...
\end{equation}
and substituting this into Eq.(2.1) leads to the following
second-order differential equations
\begin{equation}
\chi^{''}(r)+\left[(E^2-M^2)-\frac{M+E}{Mb^2}\left(\frac
{\alpha(\alpha-1)e^{-2r/b}}{(1-e^{-r/b})^2}-\frac{Ae^{-r/b}}{1-e^{-r/b}}\right)-\frac{\lambda}{r^2}\right]\chi(r)=0,
\end{equation}
\begin{equation}
\Theta ^{''}(\theta )+cot\theta ~\Theta ^{'}(\theta) +\left[\frac
{-1}{sin^2\theta }\left(\frac{M+E}{M}(\beta^{'}+\beta
cos\theta)+m^2\right) +\lambda \right]\Theta(\theta) =0.
\end{equation}

\section{BOUND STATE SOLUTION OF THE RADIAL KLEIN-FOCK-GORDON EQUATION. }

When $l\neq 0$, the differential equation in Eq.(2.3) cannot be
solved analytically due to the centrifugal term. Therefore, we must
use a proper approximation for the centrifugal term which similar
approach was also employed previously~\cite{Jia,Greene,Ahmadov}. In this
work, we attempt to use the following improved approximation scheme
to deal with the centrifugal term
\begin{equation}
\frac {1}{r^2}\approx \frac
{1}{b^2}\left[C_0+\frac{e^{-r/b}}{(1-e^{-r/b})^2} \right],
\end{equation}
which reduces to convectional approximation scheme suggested by
Greene and Aldrich when $C_0=0$ ~\cite{Greene}. For bound states
$|E|<M$, inserting this new centrifugal term into Eq.(2.3) allows us
to obtain
\begin{equation}
\chi ''(r) +\left[ {E^2-M^2 - \frac{M+E}{{M b^2}}\left(
\frac{{\alpha (\alpha  - 1)e^{-2r/b} }}{{(1 - e^{ - r/b} )^2
}}-\frac{{Ae^{ - r/b} }}{{1 - e^{r/b} }}\right) -\frac{\lambda
}{{b^2}} }\left[ {C_o + \frac{{e^{ - r/b} }}{{(1 - e^{ - r/b} )^2
}}} \right] \right]\chi (r) = 0.
\end{equation}
Eq.(3.2) can be further written in the form
\begin{equation}
\chi^{''}(s)+\frac{\tilde{\tau}}{\sigma}\chi^{'}(s)+\frac{\tilde{\sigma}}{\sigma^2}\chi(s)=0,
\end{equation}
which is known equation of the generalized hypergeometric-type by
using the transformation $s=e^{-r/b}$. Hence we obtain
\begin{eqnarray}
\chi''(s)+\chi
'(s)\frac{1-s}{s(1-s)}+\biggl[\frac{1}{s(1-s)}\biggr]^2\biggl[-\epsilon^2(1-s)^2+
A\eta s(1-s)-\alpha \eta(\alpha -1)s^2-  \nonumber \\
(1-s)^2\lambda \biggl(C_0+\frac{s}{(1-s)^2}\biggr)\biggr]\chi(s)=0,
\end{eqnarray}
where we use the following notation for bound states
\begin{equation}
\epsilon=b\sqrt{M^2-E^2}~,~~\eta =\frac{M+E}{M}.
\end{equation}

Now, we can successfully apply NU method of definition  for
eigenvalues of energy. By comparing Eq.(3.4) with Eq.(3.3) we can
define the followings

$\tilde{\tau} (s) = 1 - s,\sigma (s) = s(1 - s) $,
\begin{equation}
\tilde{ \sigma} (s) = s^2 [ - \epsilon ^2  -A\eta -\alpha
\eta(\alpha -1) - \lambda C_0 ] + s[2\epsilon ^2  + A\eta + 2\lambda
C_0 - \lambda ] + [-\epsilon ^2  - \lambda C_0 ].
\end{equation}

If we take the following factorization
\begin{equation}
\chi (s)=\phi (s)y(s),
\end{equation}
for the appropriate function $\phi (s)$ the Eq.(3.3) takes the form
of the well known hypergeometric-type equation,
\begin{equation}
\sigma (s) y^{''} (s) + \tau (s) y^{'} (s) +\bar{\lambda} y(s)=0.
\end{equation}
The appropriate $\phi (s)$ function must satisfy the following
condition
\begin{equation}
 \frac { \phi ^{'} (s)}{\phi (s)}=\frac {\pi (s)}{\sigma (s)},
\end{equation}
where $\pi (s)$, the polynomial of degree at most one, is defined as
\begin{equation}
\pi(s)= \frac{{ \sigma' -\tilde{\tau }}}{2} \pm \sqrt {(\frac{{
\sigma' -\tilde{\tau} }}{2} )^2 -\tilde{\sigma} +k\sigma}.
\end{equation}
Finally the equation, where $y(s)$ is one of its solutions, takes
the form known as hypergeometric-type if the polynomial $\bar\sigma
(s)
=\tilde\sigma(s)+\pi^2(s)+\pi(s)[\tilde\tau(s)-\sigma^{'}(s)]+\pi^{'}(s)\sigma(s)$
is divisible by $\sigma(s)$, i.e., $\bar\sigma=\bar \lambda
\sigma(s)$.

The constant $\bar \lambda$ and polynomial $\tau (s)$ in Eq.(3.8)
defined as
\begin{equation}
\bar{\lambda} =k+\pi ^{'}
\end{equation}
and
\begin{equation}
\tau (s)=\tilde{\tau} (s) +2\pi (s),
\end{equation}
respectively. For our problem, the $\pi (s)$ function is written as
\begin{equation}
\pi(s)= \frac{{ - s}}{2} \pm \sqrt {s^2 [a - k] - s[b -k] + c}
\end{equation}
where the values of the parameters are
$$
 a = \frac{1}{4} +{\epsilon ^2 + A\eta+ \alpha\eta( \alpha-1)+ \lambda C_0},
$$
$$
 b = 2\epsilon ^2 + A\eta+ 2\lambda C_0 -\lambda,
$$
$$
 c = \epsilon ^2  + \lambda C_0~.
$$

The constant parameter $k$ can be found complying with the condition
that the discriminant of the expression under the square root is
equal to zero. Hence we obtain

\begin{equation}
k_{1,2}  = (b - 2c) \pm 2\sqrt {c^2  + c(a - b)}~.
\end{equation}

When the individual values of $k$ given in Eq.(3.14) are substituted
into Eq. (3.13), the four possible forms of $\pi(s)$ are written as
follows
\begin{equation}
\pi (s) = \frac{{ - s}}{2} \pm \left\{ \begin{array}{l}
(\sqrt c  - \sqrt {c + a - b} )s - \sqrt c \,\,\,for\,\,\, k = (b - 2c) + 2\sqrt {c^2  + c(a - b)} , \\
(\sqrt c  + \sqrt {c + a - b} )s - \sqrt c \,\,\, for\,\,\, k = (b - 2c) - 2\sqrt {c^2  + c(a - b)} . \\
\end{array} \right.
\end{equation}

According to NU method, from the four possible forms of the
polynomial $\pi(s)$, we select the one for which the function $\tau
(s)$  has the negative derivative. Other forms are not suitable
physically. Therefore, the appropriate function $\pi(s)$ and
$\tau(s)$ are

\begin{equation}
\pi(s)=\sqrt{c}-s\left[\frac{1}{2}+\sqrt{c}+\sqrt{c+a-b}\right],
\end{equation}

\begin{equation}
\tau (s) = 1 +2\sqrt c -2s \left[ 1+\sqrt{c+a-b}\right]  ,
\end{equation}
for
\begin{equation}
k  = (b - 2c) -2\sqrt {c^2  + c(a - b)}.
\end{equation}

Also by Eq.(3.11) we can define the constant $ \bar{\lambda} $ as

\begin{equation}
\bar{ \lambda}=b-2c- 2\sqrt {c^2  + c(a - b)}  - \left[\frac{1}{2} +
{\sqrt{c} + \sqrt {c + a - b} }\right].
\end{equation}

Given a nonnegative integer $n$, the hypergeometric-type equation
has a unique polynomials solution of degree $n$ if and only if

\begin{equation}
\bar{\lambda}=\bar{\lambda}_n=-n\tau'-\frac{n(n-1)}{2}\sigma '',
(n=0,1,2...) ,
\end{equation}
and $\bar{\lambda}_m\neq\bar{\lambda}_n $ for
~$m=0,1,2,...,n-1$~\cite{Area}, then it follows that ,
\\

$$
\bar{\lambda} _{n_{r}}  = b-2c- 2\sqrt {c^2  + c(a - b)}
-\left[\frac{1}{2} + {\sqrt{c} + \sqrt {c + a - b} }\right]
$$
\begin{equation}
= 2n_r\left[ {1 +\left( {\sqrt c  + \sqrt {c + a - b} } \right)}
\right] + n_r(n_r - 1).
\end{equation}
We can solve Eq.(3.21) explicitly for $c$ by using the relation
 $c=\epsilon ^2  + \lambda C_0$ which brings

\begin{equation}
\epsilon^{2}  = \left[ \frac{\lambda+1/2+\Lambda  (1 + 2n_r) +
n_r(n_r + 1)-A\eta }{ 2\Lambda + 1 + 2n_r} \right]^2  -\lambda C_0,
\end{equation}
where $\Lambda =\sqrt{1/4+\eta \alpha (\alpha -1)+\lambda }$. After
inserting  $\epsilon ^2$ into Eq.(3.5) with $\lambda =l(l+1)$ for
energy levels we find

\begin{equation}
M^2-E_{n_{r},l}^2 = \frac{1}{ b^2}\left[\left[
n_{r}+1/2+\frac{(l-n_{r})(l+n_{r}+1)-A\eta }{ 2\Lambda + 1 + 2n_{r}}
\right]^2-l(l+1) C_0\right]~.
\end{equation}

The energy levels $ E_{n_r,l}$ is determined by the energy equation
Eq.(3.23), which is rather complicated transcendental equation.

Now, using NU method we can obtain the radial eigenfunctions. After
substituting   $\pi(s)$ and $\sigma(s) $ into Eq.(3.9) and solving
first order differential equation, it is easy to obtain

\begin{equation}
\phi (s)=s^{\sqrt c}(1-s)^K,
\end{equation}
where $K=1/2+\Lambda $.

Furthermore, the other part of the wave function $y_n(s)$ is the
hypergeometric-type function whose polynomial solutions are given by
Rodrigues relation
\begin{equation}
y_{n}(s) = \frac {B_{n}}{\rho (s)} \frac{{d^{n} }}{{ds^{n} }}\left[
\sigma ^{n}(s)\rho (s) \right],
\end{equation}
 where $B_n$ is a normalizing constant and $\rho (s)$ is the weight function which is the solutions of the Pearson
 differential equation. The Pearson
 differential equation and $\rho(s)$ in our case have the form,
\begin{equation}
(\sigma \rho )^{'} =\tau \rho ,
\end{equation}

\begin{equation}
\rho(s) =(1 -s)^{2K - 1} s^{2\sqrt c }.
\end{equation}
respectively.

Substitute Eq.(3.27) into Eq.(3.25) then we get

\begin{equation}
y_{n_{r}}(s) = B_{n_{r}}(1 - s)^{1 - 2K} s^{2\sqrt c }
\frac{{d^{n_{r}} }}{{ds^{n_{r}} }}\left[ {s^{2\sqrt c  + n_{r}} (1 -
s)^{2K - 1 + n_{r}} } \right].
\end{equation}
Then by using the following definition of the Jacobi  polynomials
~\cite{Abramowitz}

\begin{equation}
P_n^{(a,b)} (s) = \frac{( - 1)^n }{n!2^n (1 - s)^a (1 +
s)^b}\frac{d^n }{ds^n }\left[ {(1 - s)^{a + n} (1 + s)^{b + n} }
\right],
\end{equation}
we can write

\begin{equation}
P_n^{(a,b)} (1-2s) = \frac{C_n}{ s^a (1 - s)^b}\frac{d^n }{ds^n
}\left[s^{a+n}(1-s)^{b+n}\right],
\end{equation}
and

\begin{equation}
\frac{d^n }{ds^n }\left[s^{a+n}(1-s)^{b+n}\right]=C_n s^a (1 - s)^b
P_n^{(a,b)} (1-2s).
\end{equation}

If we use the last equality in Eq.(3.28), we can write

\begin{equation}
y_{n_{r}}(s) = C_{n_{r}} P_{n_{r}}^{(2\sqrt{c},2K-1)} (1-2s).
\end{equation}
Substituting $\phi (s)$ and $y_{n_{r}}(s)$ into Eq.(3.7), we obtain

\begin{equation}
\chi _{n_{r}}(s)=C_{n_{r}}s^{\sqrt c}(1-s)^K
P_{n_{r}}^{(2\sqrt{c},2K-1)} (1-2s).
\end{equation}
Using the following definition of the Jacobi
polynomials~\cite{Abramowitz}

\begin{equation}
P_n^{(a,b)} (s) = \frac{{\Gamma (n + a + 1)}}{{n!\Gamma (a +
1)}}\mathop F\limits_{21} \left( { - n,a + b + n + 1,1 + a;\frac{{1
- s}}{2}} \right),
\end{equation}
we  are able to write Eq.(3.33) in terms of hypergeometric
polynomials as

\begin{equation}
  \chi_{n_{r}} (s)=C_{n_{r}}s^{\sqrt c}(1-s)^{K}\frac{\Gamma (n_{r}+2\sqrt c+1)}{n_{r}!\Gamma (2 \sqrt c+1)} \mathop F\limits_{21} \left( { - n_{r},2 \sqrt c +2K+n_{r},1 +2 \sqrt c;s}\right).
\end{equation}

The normalization constant $C_{n_{r}}$ can be found from
normalization condition

\begin{equation}
\int\limits_0^\infty |R(r)|^2r^2dr=\int\limits_0^\infty |\chi (r)|^2
dr=b\int\limits_0^1\frac{1}{s}|\chi (s)|^2 ds=1,
\end{equation}
by using the following integral formula~\cite{Agboola}

$$
\int\limits_0^1 {(1 - z)^{2(\delta  + 1)} z^{2\lambda  - 1} }
\left\{ {\mathop F\limits_{21} ( - n_{r},2(\delta  + \lambda  + 1) +
n_{r},2\lambda  + 1;z)} \right\}^2 dz
$$
\begin{equation}
 = \frac{{(n_{r} + \delta  + 1)n_{r}!\Gamma (n_{r} + 2\delta  + 2)\Gamma (2\lambda )\Gamma (2\lambda  + 1)}}{{(n_{r} + \delta  + \lambda  + 1)\Gamma (n_{r} + 2\lambda  + 1)\Gamma (2(\delta  + \lambda  + 1) + n_{r})}}
\end{equation}
for $ \delta  > \frac{{ - 3}}{2}$\,\,\, and\,\,\, $\lambda >0 $.
After simple calculations, we obtain normalization constant as

\begin{equation}
  C_{n_r}=\sqrt{\frac{n_{r}!2\sqrt c(n_{r}+K+\sqrt c)\Gamma (2(K+\sqrt c)+n_{r})}{b(n_{r}+K)\Gamma (n_{r}+2\sqrt c+1)\Gamma (n_{r}+2K)} }.
\end{equation}

\section{\bf Solution of Azimuthal Angle-Dependent Part of the Klein-Fock-Gordon equation}\label{ar}

We may also derive the eigenvalues and eigenvectors of the azimuthal
angle dependent part of the KFG equation in Eq.(2.4) by using NU
method. Introducing a new variable $x=cos\theta $, Eq.(2.4) is
brought to the form
\begin{equation}
  \Theta ''(x)-\frac{2x}{1-x^2}\Theta '(x)+\frac{1}{(1-x^2)^2}\left[\lambda (1-x^2)-m^2-\eta (\beta^\prime +\beta x)\right]\Theta (x)=0.
\end{equation}

After the comparison of Eq.(4.1) with Eq.(3.3) we have

\begin{equation}
  \tilde{ \tau}(x)=-2x~~,\sigma (x)=1-x^2~~,\tilde{ \sigma}(x)=-\lambda x^2-\eta \beta x+(\lambda -m^2-\eta \beta^\prime ).
\end{equation}
In the NU method the new function $\pi (x)$ is calculated for
angle-dependent part as

\begin{equation}
\pi (x)=\pm \sqrt {x^2(\lambda -k)+\eta \beta x -(\lambda -\eta
\beta^\prime -m^2-k)}.
\end{equation}
The constant parameter $k$ can be determined as
\begin{equation}
k_{1,2}=\frac{2\lambda -m^2-\eta \beta^\prime }{2}\pm \frac{u}{2} ,
\end{equation}
where $u=\sqrt {(m^2+\eta \beta^\prime )^2-\eta^2\beta ^2}$. The
appropriate function $\pi (x)$ and parameter $k$ are
\begin{equation}
\pi (x ) =  - \left[ x \sqrt {\frac{m^2  + \eta \beta^\prime  +
u}{2}} + \sqrt {\frac{m^2  + \eta \beta^\prime  - u}{2}} \right],
\end{equation}

\begin{equation}
k = \frac{2\lambda  - m^2  - \eta \beta^\prime }{2} - \frac{u}{2}.
\end{equation}

The following track in this selection is to achieve the condition
$\tau'<0$ . Therefore  $\tau(x)$ becomes

\begin{equation}
\tau (x) =  - 2x\left[ {1 +\sqrt {\frac{{m^2 + \eta \beta^\prime  +
u}}{2}} } \right] - 2\sqrt {\frac{{m^2  + \eta \beta^\prime
-u}}{2}}.
\end{equation}

We can also write the values  $\bar\lambda =k+\pi'(s)$ as

\begin{equation}
 \bar \lambda =\frac{2\lambda -\eta \beta^\prime-m^2}{2}-\frac{u}{2}-\sqrt{\frac{m^2+\eta \beta^\prime +u}{2}}~,
\end{equation}
also using Eq.(3.20), then from the Eq.(4.8) we can obtain

\begin{equation}
\bar \lambda_{N} =\frac{2\lambda -\eta
\beta^\prime-m^2}{2}-\frac{u}{2}-\sqrt{\frac{m^2+\eta \beta^\prime
+u}{2}}=2N\left[1+\sqrt{\frac{m^2+\eta \beta^\prime
+u}{2}}\right]+N(N-1).
\end{equation}
In order to obtain unknown $\lambda $ we can solve Eq.(4.9)
explicitly for $\lambda =l(l+1)$
\begin{equation}
\lambda -\zeta ^2-\zeta =2N(1+\zeta )+N(N-1),
\end{equation}
where $\zeta =\sqrt{\frac{m^2+\eta \beta^\prime +u}{2}},$ and
\begin{equation}
\lambda =\zeta ^2+\zeta +2N\zeta +N(N+1)=(N+\zeta )(N+\zeta
+1)=l(l+1) ,
\end{equation}
then
\begin{equation}
l=N+\zeta .
\end{equation}

Substitution of this result in Eq.(3.23) yields the desired energy
spectrum, in terms of $n_r$ and $N$ quantum numbers. Similarly, the
wave function of azimuthal angle dependent part of KFG equation can
be formally derived by a process to the derivation of radial part of
KFG equation. Thus using Eq.(3.9), we obtain

\begin{equation}
\phi (x)=(1-x)^{(B+C)/2} ,
\end{equation}
where $ B=\sqrt{\frac{m^2+\eta \beta^\prime
+u}{2}}~,~C=\sqrt{\frac{m^2+\eta \beta^\prime -u}{2}}~. $

On the other hand, to find a solution for $y_N(s)$ we should first
obtain the weight function $\rho(s)$. From Pearson equation, we find
weight function as

\begin{equation}
\rho (x)=(1-x)^{B+C}(1+x)^{B-C}~.
\end{equation}
Substituting $\rho(s)$ into Eq.(3.25) allows us to obtain the
polynomial $y_N (s)$ as follows
\begin{equation}
y_{N} (x) = B_N (1 -x )^{ - (B + C)} (1 + x )^{C - B} \frac{{d^N
}}{{dx^N }}\left[ {(1 - x )^{B + C + N} (1 +x )^{B - C + N} }
\right]~.
\end{equation}

From the definition of Jacobi polynomials, we can write

\begin{equation}
\frac{d^N}{dx^N}\left[(1-x)^{B+C+N}(1+x)^{B-C+N}\right]=(-1)^N 2^N
(1-x)^{B+C}(1+x)^{B-C}P_N^{(B + C,B - C)}(x)~.
\end{equation}
Substitution  of Eq.(4.16) into Eq.(4.15) and after long but
straightforward calculations we obtain the following result,
\begin{equation}
\Theta_{N} (x) = C_N (1 - x)^{(B + C)/2} (1 + x )^{(B - C)/2}
P_N^{(B + C,B - C)} (x),
\end{equation}
where $C_N$ is the normalization constant. Using orthogonality
relation of the Jacobi polynomials ~\cite{Abramowitz} the
normalization constant can be found as
\begin{equation}
C_N=\sqrt{\frac{(2N+2B+1)\Gamma (N+1)\Gamma (N+2B+1)}{2^{2B+1}\Gamma
(N+B+C+1)\Gamma (N+B-C+1)}}~.
\end{equation}

\section{\bf Conclusion}\label{dr}

In this work we have applied NU method to the calculation of the
nonzero angular momentum solutions for the KFG equation of the
Manning-Rosen plus ring-shaped like potential. For any state energy
eigenvalues can be obtained from Eq.(3.23), which is rather
complicated transcendental equation. In case $\beta=\beta'=0$, one
can obtain central potential solutions and in case $\alpha=1$ or
$\alpha=0$ gives solutions of Hulth\'en potential. We also obtain
normalized eigenfunctions in terms of orthogonal Jacobi polynomials.

\section{\bf Acknowledgments}

The work presented in this paper was completed while one of the
authors, A.I.Ahmadov, was visiting the TH Division of the CERN. He
would like to express his gratitude to the members of the TH
Division. Financial support by CERN is also gratefully acknowledged.
The author C.Aydin thanks to members of the Department of Physics at
the University of Surrey, especially to Dr. Paul Stevenson, for
giving the opportunity to conduct research in their institute.

\end{document}